# $AgBiS_2$ near-infrared absorbers for thin film tandem photovoltaics

Shiling Dong,[1,†] Yongjie Wang,[1,2,†,*] Hae-Jun Seok,[1,†] Mengyuan Huang,[1] Edward Newton,[1] Der Ying Chen,[1,3] Yixin Wang,[1] Xiaoyu Guo,[1] Keisuke Takemoto,[4] Dusan Ivanovic,[5] Avari Roy,[6] Yongge Yang,[7] Grey Christoforo,[8] Rebekah Fraser,[8] Seongrok Seo,[8] John O'Sullivan,[9] Ha Kyung Park,[10] Jack M. Woolley,[11,12] Ji-Seon Kim,[10] Ruy S. Bonilla[9], Henry J. Snaith,[8] Qing Shen,[7] Aditya Sadhanala,[6] Lydia H. Wong,[3] Robert L. Z. Hoye[1,*]

[1.] Inorganic Chemistry Laboratory, University of Oxford, South Parks Road, Oxford OX1 3QR, UK

[2.] Institute of Functional Nano & Soft Materials (FUNSOM), Jiangsu Key Laboratory of Advanced Negative Carbon Technologies, Joint International Research Laboratory of Carbon-Based Functional Materials and Devices, Soochow University, Suzhou, 215123, Jiangsu, PR China

[3.] School of Materials Science and Engineering, Nanyang Technological University, 50 Nanyang Avenue, North Spine, Block N4., Singapore 639798

[4.] Advanced Materials and Components Laboratory, Nissan Motor Co., Ltd., 1, Natsushima-cho, Yokosuka 237- 8523, Japan

[5.] Nissan Technical Centre Europe, Cranfield, MK43 0DB, UK

[6.] Centre for Nano Science and Engineering, Indian Institute of Science, Bengaluru 560012, Karnataka, India

[7.] Faculty of Informatics and Engineering, The University of Electro-Communications, 1-5-1 Chofugaoka, Chofu, Tokyo 182-8585, Japan

[8.] Clarendon Laboratory, Department of Physics, University of Oxford, Parks Rd, Oxford OX1 3PU, UK

[9.] Electronic and Interface Materials Laboratory, Department of Materials, University of Oxford, OX13PH, Oxford, UK

[10.] Chemistry Research Laboratory, University of Oxford, 12 Mansfield Road, Oxford OX1 3TA, UK

[11.] Department of Physics, University of Warwick, Coventry, CV4 7AL UK

[12.] Warwick Centre for Ultrafast Spectroscopy, University of Warwick, Coventry, UK

[†] These authors contributed equally

* email: yongjie.wang@suda.edu.cn (Y. W.), robert.hoye@chem.ox.ac.uk (R. L. Z. H.)

## Context & scale

Tandem solar cells allow efficiencies surpassing the fundamental limit of single-junction photovoltaics, which is critical for lowering the levelized cost of energy. While crystalline Si is currently the industrial state-of-the-art near-infrared (NIR) bottom cell for perovskite-based tandems, it inherently has high embedded energy and $CO_2$ equivalent ($CO_2$eq) footprint, with limited opportunities to lower its bandgap to the optimal value. Developing tandems entirely from metal-halide perovskite thin films is highly appealing because of the potential for low capital-intensity manufacturing, low $CO_2$eq footprint, and compatibility with flexible substrates for building-integrated, mobile and space applications. However, the perovskite bottom cell for absorbing NIR light, made from alloying Pb and Sn into the metal cation site, is not ideal. $Sn^{2+}$ readily oxidizes, making handling challenging, and the bandgap can only be lowered to ~1.2 eV, limiting its NIR absorption. $AgBiS_2$ overcomes these limitations. It is stable and processable in air, and has a lower bandgap of 1.0 eV, closer to the ideal value for bottom cells in double- and triple-junction tandems (0.97 eV and 0.93 eV, respectively). This work is the first to report on the use of $AgBiS_2$ as the NIR absorber in tandem photovoltaics, and we show through optical simulations that $AgBiS_2$ enables higher efficiency limits than Pb/Sn perovskites, as well as allowing the use of lower-bandgap perovskite top-cells that are closer to their radiative limits and have improved stability. Device structures are developed that enable successful practical demonstrations of four-terminal and monolithic tandems. This work opens up a new pathway for thin film tandem photovoltaics to reach high efficiencies, with a stable bottom cell that is compatible with cost-effective manufacturing in air.

## Highlights

- $AgBiS_2$ NIR cell enables calculated 47.5% triple-junction tandem efficiency limit.
- Rapid, one-step processing in air yields thick, uniform $AgBiS_2$ nanocrystal films with state-of-the-art performance.
- Unencapsulated $AgBiS_2$ PVs retain >90% of their original efficiency after 3 months of air storage.
- First practical demonstration of $AgBiS_2$ in four- and two-terminal tandems.

## Summary

$AgBiS_2$ has rapidly risen in efficiency as single-junction photovoltaics (PVs), but its potential for tandem PVs is unexplored. Herein, we perform optical modeling based on the experimental absorption spectrum of $AgBiS_2$, revealing that, as the NIR sub-cell, efficiencies exceeding 47.5% are achievable in triple-junction tandems, superseding tandems using Pb/Sn perovskite NIR cells (44.2%). However, fulfilling this potential requires addressing the limited transport properties of $AgBiS_2$ nanocrystal films, which constrains film thicknesses to only 30–50 nm, severely restricting the NIR response. A solution-phase ligand exchange process is herein developed, allowing one-step deposition of passivated $AgBiS_2$ nanocrystals that maintain efficient charge-carrier extraction despite an order of magnitude increase in film thickness. These films enable 24.0%-efficient four-terminal tandems with visible-light-absorbing lead-halide perovskite sub-cells, and monolithic tandems successfully demonstrating voltage addition. $AgBiS_2$ PVs have high stability, retaining >90% of the original efficiency after 3 months of storage in dry air without encapsulation. This work establishes the significant potential of $AgBiS_2$ as a stable and scalable NIR absorber for tandem PVs.

## Introduction

Single-junction photovoltaics (PVs) are constrained by the Shockley-Queisser detailed-balance limit. This motivates the development of tandem PVs that partition the solar spectrum between sub-cells absorbing in complementary parts of the solar spectrum, enabling a larger fraction of the incident irradiance to be converted into electrical energy, rather than being lost as heat.[1] Metal-halide perovskites (MHPs) have progressed from a 3.8%-efficient single-junction sensitized cell in 2009[1] to perovskite/silicon dual-junction (2J) tandems with >35% power conversion efficiency (PCE)[2] as of 2026, and are undergoing intensive efforts to deploy commercially.[3] The capital- and energy-intensive crystalline silicon (c-Si) bottom cell can be replaced with a MHP sub-cell to make perovskite/perovskite tandems (already >30% PCE),[4–6] and scaled further to make triple- (3J) and quadruple junction (4J) all-perovskite tandems.[7,8] Although III-V multijunction cells retain the overall efficiency records,[9] perovskite-based tandems combine high performance with low-temperature thin-film manufacturing processes that could have lower capital intensity and carbon dioxide equivalent ($CO_2$eq) footprint than tandems using c-Si or III-V materials.[10–12] Further, all-thin film tandems benefit from compatibility with flexible substrates, which are advantageous for building-integrated applications, or as PVs integrated with the curved surfaces of electric vehicles. Whilst rapid strides have been made with wide-bandgap absorbers, the near-infrared (NIR) sub-cell in all-thin film tandems currently faces several limitations.[13,14]

The best-established NIR absorber in all-perovskite tandems is a MHP with a mixture of $Pb^{2+}$ and $Sn^{2+}$ cations in the B-site. Through bandgap bowing, this alloying can lower the bandgap towards 1.2 eV.[15,16] However, the optimal bandgap for the bottom cell in 2J and 3J tandems are approximately 0.97 eV and 0.93 eV, projecting to radiative limits in PCE of 44% and 50.7%, respectively, assuming ideal step-function absorption for each sub-cell absorber.[17] The sub-optimal bandgap of Pb/Sn perovskites only allows it to capture ~46% of the total irradiance available in the near-infrared (NIR) part of the solar spectrum (700–2500 nm wavelength).[18] Furthermore, oxidation of $Sn^{2+}$ to $Sn^{4+}$ promotes self-doping, defect formation and rapid performance loss,[19,20] necessitating careful preventative measures, such as processing in an inert glovebox environment to prevent degradation of the Pb/Sn perovskite sub-cell.

Currently, the alternatives to Pb/Sn perovskites for the NIR cell include the mature c-Si, which has been used in both 2J and 3J tandems with MHP top-cells.[21,22] Whilst c-Si is robust and stable, it has an indirect bandgap of 1.12 eV that cannot readily be reduced. c-Si is also tied to wafer-based, high-temperature, highly energy intensive processing.[10,23] $Cu(In,Ga)Se_2$ (CIGS) is a more natural partner to MHPs because its composition can be tuned to achieve a 1.0 eV bandgap. However, the multi-$TW_p$-scale deployment of CIGS is constrained by indium and gallium use, and by the need for tight control over a complex multinary phase space to suppress secondary phases that cause efficiency losses, which makes large-scale manufacturing challenging.[24] $Sb_2(S,Se)_3$ avoids the use of such scarce and expensive elements and has achieved encouraging photovoltaic performance.[25,26] However, its bandgap of 1.1–1.2 eV remains wider than the optimal value for the NIR sub-cell in multijunction tandems, limiting the fraction of the NIR spectrum that can be harvested.[27] $Cu_2ZnSn(S,Se)_4$ (CZTS) is composed of earth-abundant elements, with a bandgap tunable down to 1.0 eV. Even though recent record efficiency devices have significantly suppressed bandgap fluctuations, device performance still remains constrained by $V_{OC}$ deficits due to deep defects and non-radiative recombination,[28] stemming from Cu-Zn antisite disorder, electrostatic potential fluctuations, and band-tail states. CZTS faces the same challenges as CIGS with a narrow processing window, which makes the suppression of secondary-phases difficult.[29] NIR-absorbing organic bottom cells are appealing because they are solution-processable, flexible, and have broad molecular tunability. But the most efficient narrow-gap compositions often come with high synthetic complexity, and tandem performance is still limited by photocurrent and stability trade-offs.[30] It therefore remains a central materials challenge to develop a stable absorber with a strong NIR response, bandgap close to 1.0 eV or lower, and which is compatible with scalable thin-film processing.

Silver bismuth sulfide ($AgBiS_2$) offers this combination of properties: its elements are compliant with the restriction of hazardous substances directive (RoHS), and the material is stable in air, strongly absorbing in the NIR and processable from nanocrystal inks at low temperature.[31–33] Its ternary composition may also offer a simpler route to phase-pure absorbers than the more compositionally-complex multinary chalcogenides. In particular, the high configurational entropy of the disordered rock-salt phase of $AgBiS_2$ favors the solid solution of cations in the octahedral site, causing nanocrystals to naturally adopt this cubic structure rather than the lower-symmetry cation-ordered phase.[34,35] Further, owing to its high absorption coefficient, film thicknesses on the order of 500 nm are adequate, such that the materials intensity of Ag and Bi can be well below that of c-Si single-junction solar cells (Fig. S1, SI). $AgBiS_2$ nanocrystal films were first demonstrated as a solution-processable PV absorber a decade ago.[31] Conventional hot-injection synthesis is estimated to cost €11.6 $m^{-2}$ for a 40 nm thick film modeled under a pre-industrial process, whilst greener synthesis and lower thermal budget can reduce the cost to €4.5 $m^{-2}$.[33] This is substantially lower than reported cost estimates for PbS quantum dots (US$29–160 $m^{-2}$) and $CsPbI_3$ quantum dots (US$140–160 $m^{-2}$),[36] suggesting promising commercial viability of $AgBiS_2$. Engineering of homogeneous cation disorder in the $AgBiS_2$ nanocrystal film increases its absorption coefficient to above $10^5$ $cm^{-1}$ across the visible and NIR wavelength range (up to 1000 nm).[32] This enables an ultrathin layer of only 30–40 nm thickness to sufficiently absorb virtually all incident visible photons, although a larger thickness is still needed for NIR light. In contrast, Pb/Sn perovskites have weaker band-edge absorption ($10^4$–$10^5$ $cm^{-1}$), and typically require thicknesses on the order of hundreds of nm to a micron for adequate light absorption.[20,24]

The native long-chain organic ligands on $AgBiS_2$ nanocrystals can readily be exchanged for shorter ligands, including halide ions and thiol-containing molecules,[37] which attach to undercoordinated surface sites and passivate defects that otherwise promote non-radiative recombination. Such surface-engineering strategies have enabled certified single-junction PCEs >10%,[38] making $AgBiS_2$ the most efficient Bi-based solar absorber. So far, the reported optical bandgaps vary markedly, spanning approximately 0.8 eV in thick powder samples to 1.00–1.32 eV in quantum dot films.[39,40] However, much of this spread is due to inaccurate characterization methods; the wider bandgap values reported for nanocrystal films are largely attributable to artefacts from Tauc analysis of thin absorbing films with optical interference effects from the substrate. In this work, we establish through careful transmittance and reflectance, photoluminescence (PL), and photothermal deflection spectroscopy (PDS) measurements that the bandgap of $AgBiS_2$ NCs is 1.0 eV, and compare with the PV bandgap obtained from the inflection point of the external quantum efficiency (EQE) onset for different absorber thicknesses. This bandgap is very close to the thermodynamically ideal value for a bottom cell in multi-junction architectures. Although $AgBiS_2$ has been proposed as a candidate material for tandem PVs in recent patent applications,[41] there is, as yet, no published experimental demonstration of $AgBiS_2$ as a bottom cell in multijunction PVs, and no published tandem device performance parameters. A key reason is that the highest-performing $AgBiS_2$ devices employ ultrathin (~35 nm) absorbers due to the limited diffusion length in these materials[31]. Although single-junction devices have the $AgBiS_2$ absorber in an optical cavity by using an Ag top electrode, NIR sub-cells in tandems would not benefit from this, and these ultrathin $AgBiS_2$ films would still have an inadequate response in the NIR, where absorption coefficients are lower than in the visible wavelength range. In addition, much of the work on $AgBiS_2$ PVs has been on nanocrystals deposited to form films layer-by-layer with solid-state ligand exchange (SSLE), which makes the preparation of films >50 nm thickness highly time-consuming and impractical for large-scale, high-throughput manufacturing.

Here, we present the first experimental demonstration of perovskite/$AgBiS_2$ tandem solar cells and benchmark these against the state-of-the-art NIR absorber for perovskite-based all-thin film tandems, which are the Pb/Sn perovskites. First, we use the measured absorption profiles of these materials to perform optical simulations to quantify the improvement in NIR response that could be achieved with $AgBiS_2$ over Pb/Sn perovskites, and the potential improvement in the efficiency of 2J and 3J tandems with metal-halide perovskite top cells. Next, we develop a synthesis route for preparing $AgBiS_2$ films with >100 nm thickness that is simpler, and with shorter processing times than conventional layer-by-layer methods, but still with strong surface passivation. We examine whether this synthesis method enables adequate charge-carrier transport lengths to support thick absorbers with sufficient NIR responses for demonstrations in four-terminal tandems, and whether devices can be scaled up to the areas required for monolithic tandem demonstrations. Finally, we develop a recombination contact to enable voltage addition in monolithic tandems and protect the $AgBiS_2$ sub-cell from damage when preparing the perovskite top-cell, and also examine whether $AgBiS_2$ cells are more stable in air than Pb/Sn perovskites. Overall, this work quantifies how high the gain could be in switching from Pb/Sn perovskite NIR sub-cells to $AgBiS_2$ in terms of efficiency, stability and processability, and creates the building blocks necessary to develop practical $AgBiS_2$-based tandems that could ultimately fulfil their high potential.

# Results

## Optical simulations of perovskite/$AgBiS_2$ tandems

We begin by establishing the thermodynamic case for a 1.0 eV $AgBiS_2$ bottom cell paired with wide-bandgap perovskite top cells in 2J and 3J tandems. Detailed-balance calculations of idealized absorbers with step-function absorption edges under the AM1.5G spectrum (Fig S3, SI) reveal that reducing the bottom-cell bandgap from 1.20–1.25 eV (for mixed Pb/Sn perovskites) to 1.0 eV enables an efficiency gain across all tandem architectures. For the four-terminal (4T) geometry, the radiative-limit efficiency map (**Figure 1A**) shows that a 1.0 eV bottom cell (with an ideal step-function absorption) paired with a 1.7 eV top cell approaches a theoretical maximum of 45.6% PCE, compared to 43.7% PCE when using a 1.2 eV bottom cell. In the two-terminal (2T) current-matched geometry (**Figure 1B,** Fig S3, SI), the optimal top-cell bandgap converges to 1.65-1.70 eV as the 1.0 eV bottom-cell approaches 1 μm thickness, with a projected 2T efficiency limit of 45.2%. In a monolithic 3J tandem architecture (*i.e.*, with current matching between sub-cells; **Figure 1C**), the optimal bandgap combination is 1.9 eV, 1.35 eV and 0.9 eV for a peak PCE of 50.7%; the bandgap of $AgBiS_2$ much more closely matches the ideal 0.9 eV bottom-cell bandgap value than Pb/Sn perovskites.[38]

Moving beyond efficiency limit calculations based on idealized step-function absorption spectra, we use the experimentally-derived absorption coefficients of $AgBiS_2$ and Pb/Sn perovskites (Fig S4, SI) to calculate their ideal external quantum efficiencies (EQEs), as shown in **Figure 1D**. An 850 nm film thickness was used in these calculations to match the thicknesses used for the Pb/Sn perovskite sub-cell in Ref [42]. The $AgBiS_2$ absorber exhibits a clearly enhanced NIR response, translating directly into a higher photocurrent beyond 1000 nm wavelength. We also overlay the experimentally-measured EQE of a state-of-the-art Pb/Sn perovskite solar cell that has a 850 nm thick absorber layer,[42] which shows excellent agreement with the calculated Pb/Sn EQE at the absorption edge (**Figure 1D**), indicating that the optical properties that we are using in our simulation are realistic. **Figures 1E** and Fig S5 (SI) compare the simulated 3J tandem performance with a 850 nm $AgBiS_2$ bottom cell versus a 850 nm Pb/Sn perovskite bottom cell, respectively. A 3J perovskite/perovskite/$AgBiS_2$ stack with a 1.99 eV top cell and a 1.48 eV middle cell is projected to have a PCE of 47.5%. The maximum projected 3J efficiency with a Pb/Sn perovskite bottom cell is 44.2%, and achieving this requires a top cell with a bandgap of 2.13 eV and a middle cell of 1.65 eV. So far, very-wide-bandgap perovskites (~2.0 eV bandgap) have been increasingly developed and successfully implemented as top subcells,[43,44] but the challenge to develop high-performance perovskites with wide band gaps increases as the Br-content is increased.[45] This means that $AgBiS_2$-based 3J tandems not only have a higher efficiency limit if we can achieve high efficiency with very thick absorber layers, but they are also more practically achievable with the wide-bandgap perovskites currently available. **Figures 1F** shows the maximum calculated 3J tandem PCE as a function of bottom-cell absorber thickness over the range of 10–1000 nm, where the bandgaps of the visible sub-cells are reoptimized at each thickness. The PCE advantage of $AgBiS_2$ is more pronounced than Pb/Sn perovskites for thinner NIR absorbers. However, realizing this advantage requires the $AgBiS_2$ absorber to be substantially thicker than the 35–40 nm layers conventionally employed in high-efficiency single-junction $AgBiS_2$ PVs. Our optical simulations in Fig S6 (SI) show the maximum projected 3J PCE increases from only 32.9% with a 35 nm $AgBiS_2$ bottom cell to 42.4% at 150 nm and 45.4% at 340 nm (the maximum crack-free thickness achieved by our one-step deposition process, as detailed in the next section). Thus, increasing $AgBiS_2$ thickness to at least 210 nm (see dashed line in Figure 1F), and maintaining efficient charge-carrier extraction, is necessary to achieve 3J tandems that could surpass the best tandems theoretically possible with Pb/Sn perovskites.

Overall, these optical simulations show that $AgBiS_2$-based tandems have the potential to deliver higher efficiencies with thinner absorber layers for the bottom cell than Pb/Sn perovskites, despite having a shallower absorption onset. The use of the lower-bandgap $AgBiS_2$ as the bottom cell permits the use of narrower-bandgap perovskite top cells with lower Br content that are less susceptible to halide-ion migration and light-induced phase segregation.[46] These factors strongly motivate the practical development of $AgBiS_2$ as a bottom cell in tandems.

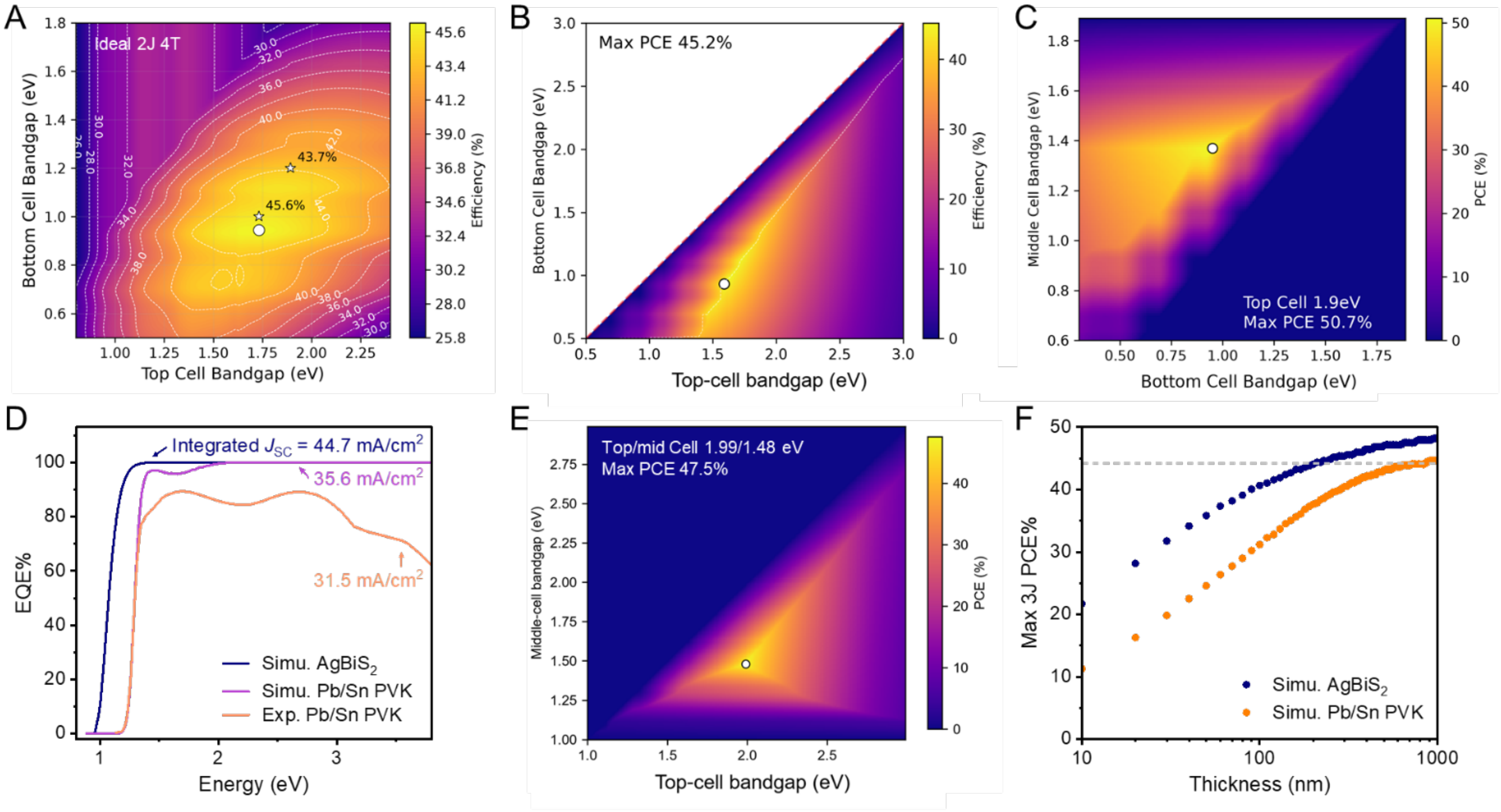

**Figure 1**. **Optical simulations of double and triple-junction tandem photovoltaics.**

In parts (A) to (C), calculations are made using step functions for the absorption profiles of the sub-cells, whereas in parts (E) and (F), calculations are made using the experimental absorption profiles of $AgBiS_2$ and Pb/Sn perovskites, while the top and mid cells have step function absorption profiles. This is to ensure that the shallower absorption onset of $AgBiS_2$ than Pb/Sn perovskites is fully accounted for.

**(A)** Detailed-balance efficiency calculated for a four-terminal tandem with different combinations of top- and bottom-cell bandgaps under AM 1.5G irradiation. The white dot marks the 1.7 eV/1.0 eV pairing, yielding a near-optimal efficiency of 45.6%. By contrast, using a 1.2 eV bandgap bottom cell pairing with 1.9 eV top cell results in a theoretical maximum efficiency of 43.7%.

**(B)** Detailed-balance efficiency calculated for a current-matched two-terminal tandem with different combinations of bandgaps for the two sub-cells. The optimal top-cell bandgap converges to ~1.7 eV for a 1.0 eV bottom cell, approaching a 45.2% 2T efficiency limit.

**(C)** Calculated maximum PCE of triple-junction solar cells with an optimal 1.9 eV bandgap for the top-cell. Refer to Fig S1, SI for the computational mapping to determine that this is the global optimal top-cell bandgap. The optimal bottom cell bandgap is 0.9 eV, which is close to the bandgap of $AgBiS_2$.
**(D)** Ideal EQE of an $AgBiS_2$ and Pb/Sn perovskite bottom cell calculated based on their measured absorption coefficients. The calculated EQE and integrated $J_{SC}$ are compared with the measured EQE and $J_{SC}$ of a state-of-the-art Pb/Sn perovskite (PVK) solar cell taken from ref [42], where the Pb/Sn perovskite film is 850 nm thick. Thus, in the calculations of the EQE, we set the absorber thickness to 850 nm for both the $AgBiS_2$ and Pb/Sn perovskite.
**(E)** Projected efficiency for triple-junction perovskite/perovskite/$AgBiS_2$ two-terminal tandems. The absorption for the top and mid perovskite sub-cells was assumed to be step functions, while the absorption coefficient for $AgBiS_2$ was from Ref. [32] . The thicknesses were set as 850 nm for the $AgBiS_2$ absorber.
**(F)** Plot of the calculated maximum efficiency of the perovskite/perovskite/$AgBiS_2$ triple-junction tandem versus the all-perovskite triple-junction tandem as a function of the thickness of the NIR sub-cell absorber. For each NIR absorber thickness, the mid- and top-cell bandgaps are re-optimized, such that the efficiencies shown are the maximum calculated values. A dashed gray line is shown as a visual aide to indicate the highest efficiency limit of the all-perovskite tandem, which has a 1000 nm thick NIR absorber. The optical data for the NIR sub-cells used in these calculations are shown in Fig S4, SI.

## Solution-phase ligand exchange with bromide passivation for $AgBiS_2$ bottom cells

To change the long-chain oleate ligands on the surface of as-synthesized $AgBiS_2$ nanocrystals to the shorter ligands required for efficient charge-carrier transport across a thin film, a ligand exchange process is required. This is commonly accomplished through solid-state ligand exchange (SSLE), which is comprised of repeated layer-by-layer deposition of the nanocrystal film. During the deposition of each layer, the nanocrystal film is treated with a solution containing the desired short-chain ligand to replace the native long-chain ligands, followed by washing to remove displaced ligands and reaction by-products. This process must be repeated to build up the desired film thickness, making SSLE lengthy and cumbersome. An alternative approach is solution-phase ligand exchange (SPLE), in which the native ligands are exchanged while the nanocrystals remain in solution. Typically, oleate-capped nanocrystals dispersed in a non-polar solvent are mixed with a polar phase containing the desired short-chain ligand. After the exchange process, the NCs are dispersed in the polar solvent, and can then be deposited to form a film in one step. So far, using SPLE to realize thick $AgBiS_2$ films has largely not been attempted due to the ultrashort carrier diffusion lengths in these materials in nanocrystal form (< 50 nm[47]).

Here, we establish a SPLE protocol that converts the oleate capped-$AgBiS_2$ nanocrystals (ABS-OA) into a hybrid surface termination comprising iodide, bromide, and chloropropane-1-thiol (CPT), denoted ABS-Br/I (**Figure 2A**). This mixed-halide strategy targets passivating the surface defects of $AgBiS_2$ nanocrystals, including undercoordinated Ag and Bi sites arising from incomplete ligand coverage or ligand loss, as well as excess Ag species associated with the Ag-rich nanocrystal surface.[32,48] Iodide and diiodoargentate anionic clusters ($[AgI_2]^-$) strongly bind to the Ag-terminated,[49] positively charged (111) facets, but weakly adsorb on nonpolar (100) facets, leaving these surfaces ligand-deficient and susceptible to defect formation and nanocrystal fusion.[50] In this work, we introduce bromide species to provide complementary passivation of sites that are less effectively stabilized by iodide alone. We confirm from X-ray diffraction (XRD) measurements that no secondary phases are detected (Fig S7, SI). Co-ligation with CPT further improves nanocrystal dispersibility through thiol binding and steric protection[51]. Thus, the control recipe is comprised of only iodide species and CPT as the ligands, denoted ABS-I. XPS quantification of the NC composition confirms that the mixed Br/I recipe (ABS-Br/I) shows three halide signals, which are 4.63 at.% Br, 4.31 at.% I, and 2.46 at.% Cl from CPT (**Figure 2B,** Fig S8 and Table S2, SI), while iodide-only (ABS-I) films have a higher halide atom-fraction of 6.32 at.% I and 4.34 at.% Cl, with no detectable Br. Notably, no detectable N 1s signal is observed in either film, indicating that nitrogen-containing methylammonium species used as the counterion for the halide species introduced are washed out and discarded during ligand exchange.

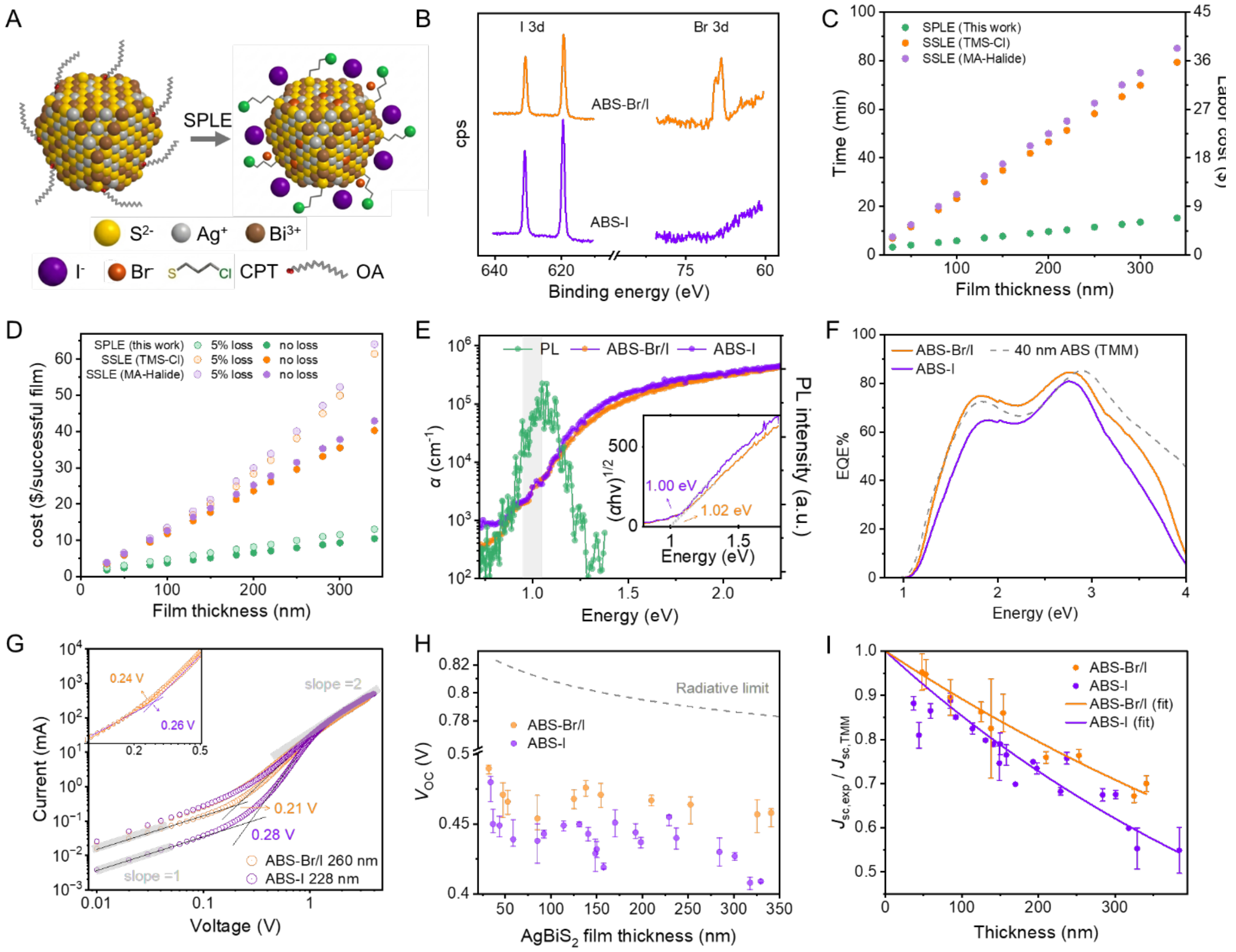

**Figure 2**. **Mixed bromide/iodide ligand engineering of $AgBiS_2$ nanocrystals.**

**(A)** Schematic of the $AgBiS_2$ nanocrystal surface chemistry after solution-phase ligand exchange: the iodide-only recipe (MAI) yields a predominantly $I^-$-terminated surface, whereas the mixed bromide/iodide recipe (MABr + MAI) produces a dual-halide surface in which the smaller $Br^-$ ions can access sub-surface Bi sites that are sterically inaccessible to $I^-$, providing deeper passivation.

**(B)** X-ray photoelectron spectroscopy of the Br 3d and I 3d core level, confirming mixed bromide/iodide incorporation in ABS-Br/I, whereas Br is absent in ABS-I.

**(C, D)** Techno-economic analysis of the labor time/cost (assumed labor wage of US\$27 $h^{-1}$) and the overall cost per film of solar absorbing layer with varied thickness up to 340 nm (as we achieved in this work). Three processes are compared, the first two involve $AgBiS_2$ deposition with the layer-by-layer solid-state ligand exchange (SSLE) method, treating the sequentially deposited layers (10 nm per layer) with methylammonium halide salt (MA-Halide)[48] or trimethylsilyl chloride (TMS-Cl)[52], versus the solution phase ligand exchange (SPLE) method used in this work. Two scenarios are evaluated in **D**: one is 'no loss', assuming 100% fabrication success rate; the other is '5% loss', assuming 95% success rate for every round of spin coating, and the failure-adjusted cost is plotted (details in Supplementary Note 1).

**(E)** Photoluminescence (PL) of ABS-OA film overlayed with absorption coefficients (α) of ABS-Br/I and ABS-I films. The absorption coefficients are obtained from transmittance and reflectance measurements of 35 nm thick films (from UV-vis). PDS measurements are scaled to the corresponding absorption coefficients in the overlapping spectral region to extend the measurement towards lower photon energies. Inset: Indirect Tauc plots derived from the combined absorption spectra, yielding optical bandgaps of 1.00 eV for ABS-Br/I and 1.02 eV for ABS-I, consistent with the PL emission peak at 1.04 eV.
**(F)** Measured external quantum efficiency (EQE) of single-junction PVs with 40 nm-thick $AgBiS_2$ absorbers compared with the calculated ideal EQE using a transfer-matrix method.
**(G)** Space-charge-limited current (SCLC) measurement of an electron-only device with either ABS-Br/I or ABS-I. The trap-filled limit (TFL) voltage is labelled for both the forward and reverse scans.
**(H)** Open-circuit voltage ($V_{OC}$) of single-junction PVs comprising ABS-Br/I or ABS-I absorbers with different thicknesses. The performance of at least four independent devices is shown for each absorber thickness, and the scatter points indicate the median and error bars indicate the full range (minimum to maximum). The thickness-dependent radiative-limit open-circuit voltages ($V_{OC}^{rad}$) is also shown, as detailed in Fig S13, SI.
**(I)** Ratio of the experimentally measured $J_{SC,exp}$ to the TMM-calculated ideal $J_{SC,\,TMM}$ (*i.e.*, which assumes 100% internal quantum efficiency) as a function of $AgBiS_2$ absorber thickness for ABS-I and ABS-Br/I solar cells.

As our optical simulations show that thicker $AgBiS_2$ absorbers are required to maximize tandem performance, we performed a techno-economic analysis (TEA) comparing the processing time and cost of the present SPLE route with two representative SSLE processes for depositing $AgBiS_2$ with thicknesses ranging from 30 to 340 nm. The scope of this TEA is strictly confined to laboratory-scale absorber manufacturing, including nanocrystal synthesis, purification, ligand exchange, and film deposition, while excluding other layers in the solar cell, solvent recovery, and bulk-purchasing discounts. To ensure an unbiased baseline, all three routes assume a single standardized synthesis batch yielding 0.9 g of $AgBiS_2$ nanocrystals, with all available ink converted into the maximum number of complete films on 1.44 $cm^2$ substrates using spin coating. The results are shown in **Figure 2C** and **2D,** with full details given in Supplementary Note 1. SSLE requires the absorber to be prepared through sequential deposition and ligand-exchange cycles, with individual layers typically limited to about 10 nm (two to three nanocrystal monolayers) to ensure effective ligand exchange throughout the film and uniform halide coverage of the nanocrystal surfaces.[48,52] Consequently, the number of coating, exchange, rinsing and drying operations increases directly with the target absorber thickness. Increasing the film thickness from 30 to 340 nm increases the calculated processing time from 7.0 to 79.3 min for the TMS-Cl SSLE route and from 7.5 to 85.2 min for the MA-halide SSLE route. In contrast, SPLE decouples ligand exchange from film formation: the nanocrystals are exchanged collectively in solutions to produce homogeneous, deposition-ready ink before coating. The film thickness can thus be readily tuned over the entire target range simply by adjusting the nanocrystal ink concentration (240-1920 mg $mL^{-1}$) and spin-coating conditions (details given in Experimental Methods). Scanning electron microscopy (SEM) measurements showed all films in the 30 to 340 nm thickness range for SPLE samples to remain compact, without cracks or pinholes (Fig S9, SI). Accordingly, the calculated processing time increases from 3.8 min (30 nm) to only 15.25 min (340 nm). This modest increase in processing time stems from the batch-wise ink production method used in the technoeconomic model; depositing a thicker film consumes more ink per unit area, resulting in a longer processing time required upstream in the nanocrystal synthesis step. At 340 nm, SPLE is therefore approximately 5.2–5.6 times faster than the two SSLE routes.

The TEA shows that this reduction in the number of unit operations in SPLE compared to SSLE translates into an increasingly pronounced economic advantage as thicker absorbers are required (**Figure 2D)**. In the idealized no-loss scenario, *i.e.,* where every spin coating step has a 100% success rate, the calculated cost per 340 nm film is about US$13.1 for SPLE, compared to US$40.3 and US$43.0 for the TMS-Cl and MA-halide SSLE routes, respectively. When the cumulative effect of processing losses is considered, *e.g.,* where each sequential step incurs an operational risk modeled at 5% loss per cycle, the calculated cost per successful 340 nm film is US$13.7 for SPLE, compared to US$61.4 and US$64.0 for the TMS-Cl and MA-halide SSLE routes, respectively. SPLE therefore costs approximately 77% less per successful film. Importantly, the one-step deposition process is more compatible with scalable, pre-metered techniques, such as blade and slot-die coating, which are widely used for large-area PV manufacturing.[53,54]

However, this single-step deposition approach does have practical limitations. We found that when the film thickness is close to approximately 400 nm, the thermal stresses induced by solvent evaporation caused the film to crack (Fig S10, SI). Consequently, the maximum thickness that maintains structural integrity in a single deposition is approximately 350 nm. But, as mentioned in the previous section, our optical simulations show that increasing $AgBiS_2$ thickness to 350 nm in a 3J tandem is already adequate to raise the radiative limit beyond that when using a Pb/Sn perovskite bottom cell (**Figure 1F**). Similarly, for the more practically relevant 2T 2J configuration that we will demonstrate in the tandem device section, increasing the thickness of $AgBiS_2$ from 35 to 350 nm increases the maximum projected PCE from 24.55% to 31.52% (Fig S11, SI). Thus, the thickness accessible through our one-step process is already sufficient to substantially increase the efficiency potential of both 2J and 3J tandems. Greater thickness could be achieved through multiple deposition steps with intermediate heat treatments, for example, reaching 850 nm after three depositions.

To determine the intrinsic optical bandgap of $AgBiS_2$, we combine transmittance and reflectance measurements obtained from UV-visible spectrophotometry (UV-vis) together with photothermal deflection spectroscopy (PDS), enabling the determination of the absorption coefficient across both the strongly absorbing above-gap and weakly absorbing near-band-edge regions (**Figure 2E**). PDS only provides the relative absorbance, whereas UV-vis measurements are affected by interference effects close to the band-edge. We therefore use UV-vis to obtain the absolute absorption coefficients in the above-gap region (from Beer-Lambert law), and linearly scale the absorbance from PDS to match the UV-vis-derived absorption coefficients in the 2.28–2.81 eV range, where absorbance from PDS and UV-vis have close spectral match. The scaled absorbance from PDS is used to estimate the absorption coefficient at photon energies below 2.3 eV, since this is in this range where interference effects start to affect the UV-vis measurements. The overall absorption coefficient spectrum is used to construct a Tauc plot for indirect-bandgap materials (**Figure 2E**, inset), and from this, the bandgaps obtained are 1.00 eV for ABS-Br/I and 1.02 eV for ABS-I, consistent with the previously reported electronic structure of the disordered rock-salt phase.[32] The PL emission peaks at 1.04 eV, which is marginally above the PDS-measured band edge, consistent with the first direct transition of $AgBiS_2$ being slightly wider than the indirect bandgap. However, the sub-bandgap absorption of ABS-Br/I is measurably lower than that of ABS-I across the 0.7-1.0 eV window, indicating that the additional bromide surface co-passivation provides a reduction in mid-gap trap-state density.

As shown in **Figure 2F**, the EQE of ABS-Br/I thin film exceeds that of ABS-I with similar absorber thicknesses of 35-40 nm, especially at longer wavelengths, and approaches the transfer-matrix-method (TMM) calculated EQE for an ideal PV with a 40 nm thick absorber. This shows that in this thin absorber regime, we have more efficient charge-carrier extraction with ABS-Br/I, which then makes the EQE measurements suitable for estimating the effective photovoltaic bandgap. We compare these measurements with the ideal EQE, which is calculated through three separate models: (1) single-pass absorption and (2) ideal double-pass absorption with the Beer-Lambert law, and (3) the transfer-matrix method (TMM). Method 3 accounts for how the absorber thickness and device optical architecture influence the spectral response of $AgBiS_2$ photovoltaics (Fig. S12, SI). The corresponding PV bandgaps, extracted from the first derivative of the calculated EQE spectra, shift towards higher energies with decreasing absorber thickness (Fig. S13, SI), reflecting the reduced absorption of lower-energy photons in thinner films.

The improvement in EQE with the mixed Br/I passivation may arise from a reduction in defect density, as suggested by the PDS measurements (**Figure 2E**). To investigate this, we performed space-charge-limited current (SCLC) measurements on electron-only $AgBiS_2$ devices (**Figure 2G**). We found that devices with $AgBiS_2$ films as thin as 30-50 nm are not suitable for SCLC because the current can exceed the compliance limit of the source meter before the space-charge-limited regime is fully resolved, and the much higher electric field across a thin film increases the risk of electrical breakdown before the required space charge regime is reached. Here, we increased the film thickness to 260±8 nm for ABS-Br/I, and 228±5 nm for ABS-I; the thickness is not increased further because otherwise the $V_{OC}$ decreases markedly, especially for ABS-I (**Figure 2H**). In the ABS-I electron-only devices, a hysteresis between the forward and reverse sweeps reflects slow trap-filling dynamics, and the trap-filled-limit (TFL) transition voltage is identified at 0.28 V and 0.26 V for the two sweep directions, yielding an average trap density of $1.36 \times 10^{15}$ cm$^{-3}$. For ABS-Br/I, the hysteresis is reduced, along with reduced TFL voltages of 0.21 V and 0.24 V in the two sweep directions, corresponding to a trap density of $1.07 \times 10^{15}$ cm$^{-3}$. These measurements therefore confirm that dual passivation lowers trap densities. Consistent with this, the $V_{OC}$ of PV devices with the same thickness of $AgBiS_2$ as used for SCLC measurements are 0.464±0.014 V for ABS-Br/I, compared to 0.455±0.006 V for ABS-I. To quantify the significance of this $V_{OC}$ improvement, we determine the radiative limit to be 0.78 V from the 1.04 eV energy of the PL peak (close to the $V_{OC}^{rad}$ calculated from the ideal EQEs in Fig S13, SI). The $V_{OC}$ deficit is therefore reduced from 0.325 V for ABS-I to 0.316 V for ABS-Br/I. Fig S14 (SI) compares the $V_{OC}$ of these devices with the $AgBiS_2$ single-junction PVs reported in literature employing SPLE or SSLE strategies. Notably, the ABS-Br/I devices achieve $V_{OC}$ comparable to those of the reported high-performing $AgBiS_2$ nanocrystal solar cells (30-50 nm film thicknesses[37,55]) even though the absorber layer is several times thicker and deposited at a higher deposition rate, which are important advantages for tandem applications.

X-ray photoelectron spectroscopy (XPS) and ultraviolet photoelectron spectroscopy (UPS) measurements provide further insights into the mechanism of surface defect passivation. Ligand exchange from ABS-OA to halide ligands induces shifts in the Bi 4f, Bi 5d and Ag core levels toward higher binding energy, indicating reduced electron density at the metal sites. The shift is larger for ABS-I (0.39 eV) than for the mixed Br/I-treated sample (0.22 eV). Detailed core-level fits and the corresponding shifts for Bi 4f, Bi 5d and Ag 3d are provided in Fig S15 and S16 (SI). The Bi 4f and Bi 5d spectra present and are fitted with Bi–S and Bi–halide components. The Bi–halide/Bi–S area ratios are 0.32 for ABS-I and 0.42 for ABS-Br/I, indicating a larger fraction of Bi in the higher-binding-energy, halide-coordinated environment for the mixed Br/I treatment, suggesting that $Br^-$ contributes more effectively to the passivation of subsurface or less accessible defects by larger $I^-$. We checked for changes in Fermi level but found the valence band to Fermi level offset to be similar for all samples (Fig S17, SI), indicating that the shifts in core levels are not due to changes in doping level. The vacuum level, however, does change after ligand exchange, with the work function increasing from 4.01 eV for ABS-OA to 4.57 eV for ABS-I and 4.39 eV for ABS-Br/I (Table S9, SI). The pronounced work-function changes are therefore consistent with an additional vacuum-level shift associated with the modification of nanocrystal surface chemistry. Similar effects have been reported for ligand-exchanged PbS nanocrystals where ligand exchange can shift nanocrystal energy levels by up to 0.9 eV due to surface dipoles associated with ligand-nanocrystal interactions, while the bandgap remains unchanged.[56] The energy-level diagram of the solar cell is constructed, showing the band alignment of $AgBiS_2$ with charge-extracting layers like $SnO_2$ and PTAA (Fig S18, SI).

To assess how the Br co-passivation translates into device performance across a range of absorber thicknesses, we measure the single-junction PV performance and evaluate how the measured $J_{SC}$ varies relative to the TMM-calculated $J_{SC,TMM}$ for ABS-Br/I and ABS-I solar cells with the active-layer thicknesses tuned from 35 to 340 nm (**Figure 2I**). ABS-Br/I maintains consistently higher $V_{OC}$ than ABS-I across the full thickness range (**Figure 2H**) despite a gradual decrease with increasing film thickness. A similar thickness-dependent decrease is observed in the calculated radiative-limit open-circuit voltages ($V_{OC}^{rad}$) obtained using a detailed-balance approach based on single-pass absorption-derived EQE spectra (**Figure 2H**). ABS-Br/I and ABS-I solar cells show similar $J_{sc}$ values (Fig S19, SI), in agreement with the predicted $J_{sc}$ by TMM optical simulations, where the measured $J_{sc}$ shows a characteristic dip at around 80 nm thickness caused by destructive interference across the layered thin film optical cavity. We note that this dip would only occur for the specific device layout we have for these single junction devices ($ITO/SnO_2/AgBiS_2/PTAA/MoO_x/Ag$), and would not expect such strong optical cavity effects when used in tandem devices. Both single-junction $AgBiS_2$ nanocrystal devices exhibit negligible optical reflection losses (< 1%) across the investigated thickness range (Fig S20, SI).

Table S10 (SI) summarizes the performance of representative devices from this work. In the single-junction $AgBiS_2$ solar cell configuration, our champion device employs a 35±3 nm-thick ABS-Br/I absorber and exhibits a $V_{OC}$ of 0.503 V, $J_{SC}$ of 26.31 mA cm$^{-2}$, and PCE of 8.98%, similar to the current certified record efficiency of single-junction $AgBiS_2$.[32] The measured $J_{SC,exp}$ to calculated $J_{SC,TMM}$ ratio approaches 95%, indicating nearly complete charge-carrier collection throughout the active layer. Such efficient carrier extraction is enabled by the exceptionally high absorption coefficient of $AgBiS_2$ (>$10^5$ cm$^{-1}$ across the 400–1000 nm spectral region[32,50] ), together with optical confinement and multiple internal reflections within the device stack. Our second-best device is based on a 150±4 nm ABS-Br/I film, showing a $V_{OC}$ of 0.482 V, $J_{SC}$ of 26.52 mAcm$^{-2}$ and PCE of 7.8%, where the higher $J_{SC}$ is ascribed to enhanced NIR light harvesting, consistent with optical simulations. As the film thickness increases beyond the carrier diffusion length of the $AgBiS_2$ nanocrystal film (previously reported as approximately 30-50 nm for $AgBiS_2$ with MPA ligands [32]), the $J_{SC,exp}/J_{SC,TMM}$ of both ligand recipes declines. Nevertheless, the consistently higher ratios observed for ABS-Br/I indicate that bromide incorporation alleviates thickness-dependent collection losses, enabling efficient operation at substantially greater absorber thicknesses. At a thickness of 340 nm, ABS-Br/I device presents a PCE of 6.05%, whereas iodide-treated counterparts are limited to approximately 4% PCE. This is mainly because ABS-I devices exhibit a pronounced decline in $V_{OC}$ once the absorber thickness exceeds approximately 250 nm (**Figure 2G**), whereas ABS-Br/I devices maintain a smaller $V_{OC}$ deficit across the thickness range investigated, suggesting that bromide incorporation suppresses thickness-dependent non-radiative recombination losses.

## Demonstration of perovskite/$AgBiS_2$ tandems

Having now developed a process for rapidly depositing thick $AgBiS_2$ absorbers with high single-junction performance, we now focus on making the first demonstration of perovskite/$AgBiS_2$ tandems. We start with 4T tandems by pairing an $AgBiS_2$ bottom cell with a semitransparent perovskite top cell (**Figure 3A**). The 4T configuration serves as a natural starting platform to validate spectral complementarity without the added constraint of strict sub-cell current matching. The combined tandem efficiency is the sum of the individual sub-cell efficiencies under standard one-sun illumination.

We evaluate two semitransparent top-cell variants: a mid-bandgap (MBG, 1.55 eV) perovskite (with a nominal composition of $Cs_{0.05}FA_{0.85}MA_{0.1}PbI_3$; where FA is formamidinium) device with a single-junction PCE of 21.05%, and a wide-bandgap (WBG, 1.68 eV) perovskite (with a nominal composition of $Cs_{0.17}FA_{0.83}Pb(I_{0.6}Br_{0.4})_3$) device with a lower PCE of 16.25% but whose wider optical gap provides better spectral complementarity to the $AgBiS_2$ bottom cell. Both perovskite devices have a sputter-deposited $MgF_2$ antireflection layer over the transparent top electrode. The $AgBiS_2$ bottom cell is operated in a bottom-illumination geometry, receiving photons transmitted through the semi-transparent perovskite top cell. A 4T efficiency of 23.94% was achieved with 150 nm $AgBiS_2$ coupled with MBG top cell (21.05% top + 2.89% bottom), and 19.49% with the WBG top cell (16.25% top + 3.24% bottom). Current-voltage characteristics of the individual sub-cells and the spectrally filtered $AgBiS_2$ bottom cell are shown in **Figure 3B**, with the inset of Fig S21 (SI) showing a photograph of how these two sub-cells were placed together. EQE of the 150 nm-thick ABS-Br/I bottom cell, measured beneath each top cell, extends from the visible to beyond 1200 nm and remains reasonably close to the TMM-calculated reference EQE at the visible range. Some current deficit is observed at the longest wavelengths, attributable to photogenerated charge-carriers produced at depths exceeding the carrier diffusion length within the 150 nm $AgBiS_2$ layer. We investigated the PCEs of 4T tandems as a function of $AgBiS_2$ absorber thickness (32–340 nm) using MBG and WBG semitransparent top-cells respectively (**Figure 3C**). 150 nm $AgBiS_2$ gives the maximum PCE due to a balance of optical cavity effects for enhanced light capture, stronger NIR absorption compared to 35 nm thinner absorbers, along with adequate carrier collection. These results establish the viability of $AgBiS_2$ in a tandem geometry and underscore how the thicker films with mixed-halide passivation can deliver competitive PCEs. In particular, it can be seen from **Figure 3F** that this first report of a $AgBiS_2$/perovskite 4T tandem surpasses the first report of 4T perovskite/Si and perovskite/perovskite tandems.

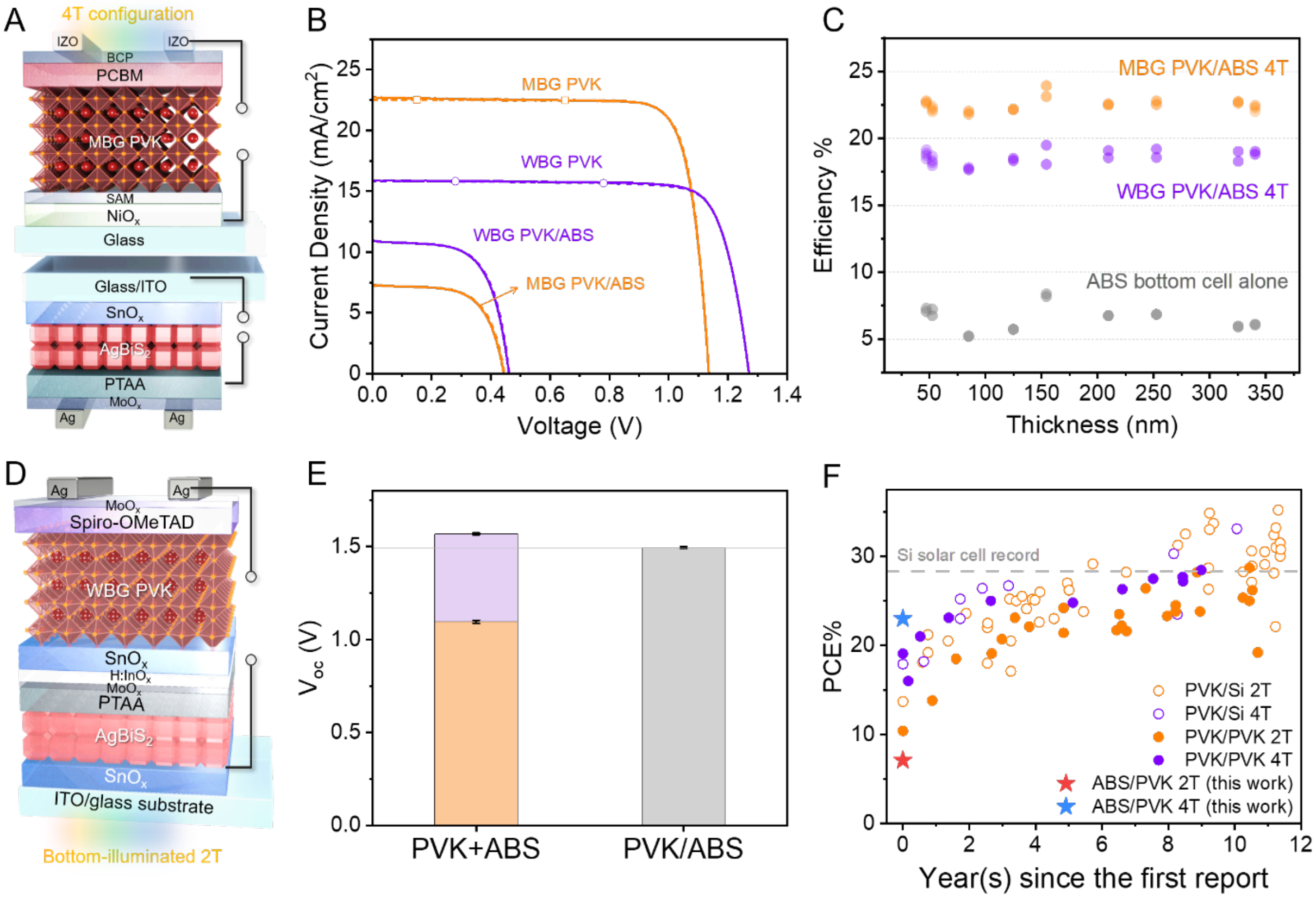

**Figure 3. Demonstration of the first perovskite/$AgBiS_2$ tandems.**

**(A)** Device schematic of the mechanically stacked four-terminal configuration.

**(B)** *J-V* curves under simulated AM1.5G illumination of the MBG semitransparent perovskite top cell, the WBG semitransparent perovskite top cell, and the 150 nm $AgBiS_2$ bottom cell measured under the spectra filtered through the MBG and WBG top cells. The forward scans are plotted in solid lines while the reverse scans are as dashed lines.

**(C)** Combined 4T PCE as a function of $AgBiS_2$ absorber thickness (32 to 340 nm) for MBG/$AgBiS_2$ and WBG/$AgBiS_2$ pairings.

**(D)** Device schematic of the monolithic two-terminal (0.04 $cm^2$) tandem stack: ITO/$SnO_2$/$AgBiS_2$/PTAA/$MoO_x$/H:$In_2O_3$/$SnO_2$/perovskite/Spiro-OMeTAD/$MoO_x$/Ag. Here, for the first demonstration, these monolithic tandems are illuminated from the bottom through the $AgBiS_2$ sub-cell

**(E)** $V_{OC}$ comparison of the individual perovskite and $AgBiS_2$ single-junction sub-cells versus the 2T monolithic tandem.

**(F)** Efficiency trajectory of perovskite/perovskite and perovskite/Si tandems in both 2T and 4T architectures plotted as a function of years since the first reported publication. The chosen PCEs are from devices with area more than 0.1 $cm^2$, apart from the publications reporting the first operational 2T and 4T perovskite/silicon and all-perovskite tandem solar cells.[15,57–59] The present perovskite/$AgBiS_2$ results (0.25 $cm^2$ device area) are labeled by star on the tandem roadmap. Data in (F) are obtained from ref [15,53,57,58,60–69]. The dashed grey line labels the current certified record 28.29% PCE for single-junction Si cell (LONGi, 24 September 2026[70]).

Monolithic integration imposes more stringent requirements on the device stack than the 4T geometry, and our primary objective here is to demonstrate the experimental feasibility of this 2T concept, where $V_{OC}$ addition proves the efficacy of our proposed tandem structure and processing routes. A particular challenge is in developing a recombination contact that not only enables voltage addition, but also protects the bottom cell from damage when processing another sub-cell on top. Proving the feasibility of combining $AgBiS_2$ with another sub-cell is the key building block that opens up future opportunities to create more complex 3J and 4J tandems.

For the initial demonstration of a 2J monolithic tandem, we employed a device architecture comprising $ITO/SnO_2/AgBiS_2/PTAA/MoO_x/H:In_2O_3/SnO_2$/perovskite/Spiro-OMeTAD/$MoO_x$/Ag (cross-sectional SEM image in Fig S22, SI). The perovskite has the nominal chemical formula of $Cs_{0.05}FA_{0.85}MA_{0.10}PbI_3$, with a bandgap of 1.55 eV. In this first demonstration, we use thin $AgBiS_2$ absorbers (35 nm thickness), and we illuminate through the $AgBiS_2$ bottom cell to solely focus on determining whether we could achieve voltage addition; 2T tandems will ultimately have to be illuminated through the perovskite top-cell (see next section). We introduce $H:In_2O_3$, grown by atmospheric pressure chemical vapor deposition (AP-CVD) as the recombination contact. This is a transparent conducting electrode that has a high average NIR transmittance of 89%, and is highly *n*-type doped, enabling recombination of excess charge-carriers from each sub-cell.[71] Growth by AP-CVD enables a shorter processing time than atomic layer deposition, but with similar electronic, optical and morphological properties, which are important for minimizing thermally-induced damage to the bottom-cell. Atomic force microscopy shows that the 15 nm $H:In_2O_3$ films deposited at 100 °C on glass or $MoO_x/AgBiS_2$ exhibit smooth and uniform surfaces, with RMS roughness values of 1.37 nm (on glass) and 1.60 nm (on $MoO_x/AgBiS_2$), as shown in Fig S23 (SI). Previously we shown that $H:In_2O_3$ grown on glass with a thickness of 710±50 nm has a mobility of 190±30 $cm^2\ V^{-1}\ s^{-1}$, with a low sheet resistance of 7.20±0.01 $\Omega\ \square^{-1}$.[71] In this work, the same deposition recipe was used to prepare the substantially thinner $H:In_2O_3$ recombination contact.

To test this recombination contact, we initially fabricate devices over small area (0.04 $cm^2$) on conventional ITO/glass substrates and illuminate from the $AgBiS_2$ bottom cell, which uses a 35 nm thick ABS-Br/I film. The measured $V_{OC}$ of the 2T tandem device is 1.49 V, which is 95.5% of the sum of the $V_{OC}$ of the $AgBiS_2$ bottom cell (0.47 V) and perovskite top-cell (1.09 V after filtering with 35 nm $AgBiS_2$; the unfiltered $V_{OC}$ is 1.10 V), as shown in **Figure 3E** and Fig S24 (SI). The $J_{SC}$ of this bottom-illuminated 2T tandem is 9.4 mA $cm^{-2}$, and together with a fill factor of 0.57, the device delivers an overall PCE of 8%. The near-additive photovoltage therefore demonstrates effective electrical coupling of the $AgBiS_2$ and perovskite sub cells through $PTAA/MoO_x/H:In_2O_3/SnO_2$ layers, establishing the feasibility of monolithic integration.

## Scalability and stability of $AgBiS_2$ devices

Thus far, $AgBiS_2$ has been a novel and emerging solar absorber, and, as such, the device areas reported in the literature have been small, *i.e.*, $\leqslant$ 0.04 $cm^2$. Having successfully realized a device architecture that achieves voltage addition in monolithic tandems, our work strongly motivates maturing $AgBiS_2$ PVs to the larger device areas expected in the perovskite-based tandems field, along with developing a top-illuminated device architecture. **Figure 4A** presents the *J-V* curves for 150 nm-$AgBiS_2$ single-junction solar cells fabricated with two device areas using the ABS-Br/I ink synthesized from the same batch. The typical lab-scale device (0.04 $cm^2$ area) yields a champion PCE of 7.3%, and these were prepared on 1.2 × 1.2 $cm^2$ area substrates (ITO/glass). We scale up our one-step ABS-Br/I recipe to a 3 × 3 $cm^2$ area substrate (FTO/glass), where we have six 0.25 $cm^2$ area devices (**Figure 4A**, inset). Despite the order of magnitude increase in device area, the performance is comparable to the small devices, with a champion PCE of 6.5%, and pixel-to-pixel PCE variation on the same substrate less than 0.6%. The slight efficiency reduction upon scale-up is primarily attributable to a reduction in the measured $J_{SC}$ (Table S10, SI). The $V_{OC}$s are similar (**Figure 4A**), suggesting that the larger-area films are not limited by increased pinhole density, or a higher probability of having pinholes within the active device area. **Figure 4B** compares the EQE spectra and reveals that the EQE-integrated $J_{SC}$ values are similar, at 25.4 and 25.1 mA $cm^{-2}$ for the 0.04 and 0.25 $cm^2$ devices, respectively, indicating that the spectral response and intrinsic photocurrent generation are largely preserved upon scaling (more discussion in Supplementary Note 2, SI). Importantly, the SPLE-based ink deposition process also produces highly uniform $AgBiS_2$ films over the entire 3 × 3 $cm^2$ substrate. We performed variable angle spectroscopic ellipsometry measurements across eight spatial locations, covering both peripheral and central regions, for a $AgBiS_2$ nanocrystal film with a nominal thickness of 200 nm, deposited on a 3 × 3 $cm^2$ glass substrate (Fig S25, SI). We observe exceptionally low fit uncertainties (standard deviations 0.37-0.82 nm), no systematic thickness gradient, and the extracted film thicknesses variation below 4 nm. This uniformity highlights an important advantage of the SPLE approach for extending $AgBiS_2$ nanocrystal deposition beyond small device areas.

Having established the reproducible fabrication of up-scaled $AgBiS_2$ single-junction devices, we next develop top-illuminated 2T tandems with an active area of 0.25 $cm^2$. Cr/Au-coated glass is used as the substrate for the $AgBiS_2$ bottom cell, consisting of $SnO_2$/$AgBiS_2$/PTAA/$MoO_3$, followed by the AP-CVD H:$In_2O_3$ recombination contact developed above. The WBG perovskite top cell employs $Cs_{0.17}FA_{0.83}Pb(I_{0.6}Br_{0.4})_3$ as the absorber and $(In_2O_3)_{90}(ZnO)_{10}$ (IZO) as the transparent top electrode. The complete device architecture is shown schematically in **Figure 4C**, with illumination through the IZO side. Nine individual tandem devices are fabricated on each 3 × 3 $cm^2$ substrate, with each device 0.25 $cm^2$ in area. The $V_{OC}$ comparison of the individual sub-cells and the tandem is presented in **Figure 4D**. The $AgBiS_2$ and WBG-perovskite single-junction controls exhibit average $V_{OC}$ values of 0.44 ± 0.04 V and 1.25 ± 0.06 V, respectively, while the 0.25 $cm^2$ monolithic tandems exhibit an average $V_{OC}$ of 1.36 ± 0.07 V. The tandem therefore retains approximately 80% of the sum $V_{OC}$ of the two single-junction devices (1.69 V). The clear voltage addition demonstrates the operation of the recombination junction after integration into the scaled-up device, although the approximately 0.33 V deficit relative to the sum of the individual sub-cell voltages indicates that further work is needed to reduce interfacial and recombination losses. The *I-V* curve of the 2T tandem is presented in Fig S26 (SI). The champion 0.25 $cm^2$ 2T monolithic tandem delivers a $V_{OC}$ of 1.35 V, a $J_{SC}$ of 10.46 mA $cm^{-2}$, fill factor of 50.3%, and a PCE of 7.11%, which is comparable to the efficiencies achieved in the early development of other monolithic tandem technologies (**Figure 3F**). For comparison, the first reported monolithic perovskite/perovskite, perovskite/CIGS, and perovskite/CZT(SSe) tandem solar cells exhibited PCEs of approximately 10.4%,[69] 3.8%,[72] and 4.6%,[73] respectively. With this first demonstration of monolithic $AgBiS_2$/perovskite tandems, the main factor limiting the efficiency is the fill factor of ~50%, which is lower than the fill factors of 68% and 65% corresponding to the single-junction WBG-perovskite and $AgBiS_2$ devices, respectively. The reduced tandem fill factor therefore points to additional losses introduced upon monolithic integration in these early tandem demonstrations, potentially including increased series resistance, transport losses through the interconnection layers, or non-ideal charge recombination at the recombination contact. Further discussion regarding the optimization of top-illuminated 2T tandems is provided in Supplementary Note 2.

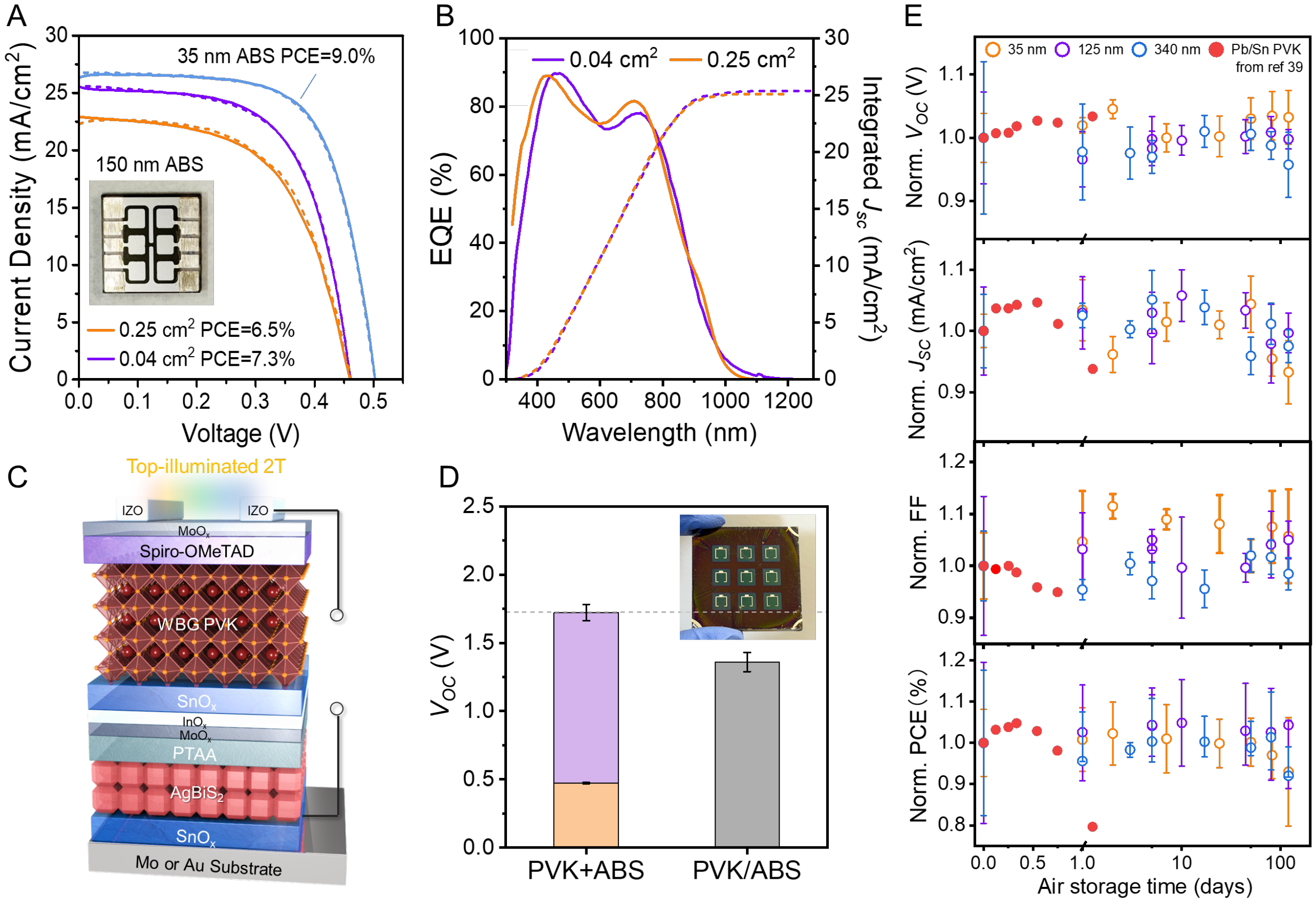

**Figure 4. Scalability and stability of $AgBiS_2$ devices.**

**(A)** *J–V* curves of $AgBiS_2$ single-junction solar cells with active areas of 0.04 cm$^2$ (PCE = 9.0% for 35 nm thickness, PCE = 7.3% for 150 nm thickness) and 0.25 cm$^2$ (PCE = 6.5% for 150 nm thickness). Inset: photograph of an array of 0.25 cm$^2$ devices on a 3 × 3 cm$^2$ area substrate. The forward scans are plotted in solid lines while the reverse scans are as dashed lines.

**(B)** EQE spectra of the 0.25 cm$^2$ device compared with the small-area (0.04 cm$^2$) device that were fabricated at the same batch, with the corresponding integrated short-circuit current densities indicated. The similar integrated current densities show that photocurrent generation is largely preserved upon increasing the device area.

**(C)** Device schematic of the large-area (0.25 cm$^2$) monolithic perovskite/$AgBiS_2$ tandem. Illumination is from the IZO side.

**(D)** $V_{OC}$ comparison of the individual sub-cells and the large area 2T monolithic tandem.

**(E)** Shelf-life stability comparison of unencapsulated $AgBiS_2$ solar cells based on ABS-Br/I film with varied thicknesses. We compare these results with a literature report of a highly passivated Pb/Sn perovskite device obtained from ref. [42]. Both devices are stored under ambient dry conditions with sunlight exposure during periodic measurements. The Pb/Sn device degrades to <80% of initial efficiency within 1.5 days, whereas $AgBiS_2$ retains >90% over three months. Data are presented as mean values ± standard deviations based on at least five measurements.

Finally, a critical bottleneck in all-perovskite tandem PVs is the instability and processing sensitivity of the narrow-bandgap mixed Pb/Sn perovskites bottom cell, which can degrade rapidly due to the spontaneous oxidation of $Sn^{2+}$ to $Sn^{4+}$. In contrast, $AgBiS_2$ does not contain oxidatively unstable ions, enabling ambient-air ink handling and scalable deposition without more complicated atmosphere control during processing, which could reduce capital and operational costs in commercial manufacturing, as well as potentially enabling higher throughputs to be achieved more easily. **Figure 4E** presents the shelf-life stability of unencapsulated $AgBiS_2$ solar cells that we prepared, which we compare against literature reports of a highly passivated mixed Pb/Sn perovskite single-junction PV device. In both cases, the devices are stored under ambient dry conditions, and periodically measured in air under 1-sun illumination, which approximately corresponds to the ISOS-D-1 protocol.[74] Despite representing a state-of-the-art, highly passivated formulation, the literature report of the Pb/Sn perovskite device degrades to below 80% of its initial efficiency within just 1.5 days.[42] By contrast, our unencapsulated $AgBiS_2$ devices retain more than 90% of their initial efficiency after three months of storage in similar conditions. Promisingly, all of the performance parameters remain within error of the original $V_{OC}$, $J_{SC}$ and fill factor values, and this is consistent across 35, 125 and 350 nm thick absorbers (**Figure 4E**). We note that this is a promising result, indicating the potential for high stability for $AgBiS_2$ solar cells, although further work is required to ensure that the solar cells are stable under combined heat and light stressing, and when deployed outdoors.

## Discussion

The historical development of perovskite-based tandems illustrates how rapidly a newly demonstrated absorber combination can progress, as shown in **Figure 3F**. It is highly promising that our work identifies the combination of materials that enables the first demonstration of 4T and 2T tandems with $AgBiS_2$ PVs, with efficiencies comparable to, or surpassing, the first reports of perovskite/Si, perovskite/perovskite and other perovskite-based thin film tandems. The challenge now is to map out how $AgBiS_2$-based 2J tandems could follow their trajectory of efficiency improvements, and be further developed into 3J tandems, where their low bandgap becomes particularly advantageous.

When comparing the measured EQE of the 340 nm-thick single-junction ABS-Br/I device, *i.e.,* the thickest $AgBiS_2$ device reported to date, with the TMM-calculated EQE, there is a pronounced deficit, especially at longer wavelengths, where low-energy photons are absorbed deeper within the film (Fig S27, SI). This shortfall may arise because the carrier diffusion length in the nanocrystal film is shorter than the absorber thickness, such that photogenerated carriers created deep in the layer recombine non-radiatively before reaching the charge-selective contacts. To test this hypothesis, we use optical pump terahertz probe (OPTP) spectroscopy, where the differential transmittance is proportional to the change in photoconductivity, and therefore depends on the charge-carrier lifetime, as well as the charge-carrier mobility (note that $AgBiS_2$ has free carriers rather than excitons[35]). OPTP measurements (Supplementary Note 3) confirm that both ABS-Br/I and ABS-I nanocrystal films still both undergo ultrafast carrier localization, with $t_{50} < 1$ ps and a Drude-Smith constant of 0.9, indicating that localization is intrinsic to these nanocrystal films (detailed parameters and analysis are given in Supplementary Note 3, Table S11, SI). Bromide passivation does not prevent carrier localization from taking place, but reduces the trap-assisted recombination rate constant $k_1$ from $2.12 \times 10^{10}$ $s^{-1}$ for ABS-I to $0.75 \times 10^{10}$ $s^{-1}$ for ABS-Br/I, consistent with the lower trap density from SCLC and PDS measurements. Thus, while Br/I co-passivation enables improved diffusion lengths, supporting the development of thicker $AgBiS_2$ absorbers in devices, carrier localization remains an inherent bottleneck.

We have very recent fundamental work showing that a pathway to overcome these carrier localization limitations in $AgBiS_2$ is to prepare bulk films with low grain boundary density and energetic disorder.[35] This led to an order of magnitude improvement in mobility up to 4.16 $cm^2\ V^{-1}s^{-1}$, with diffusion lengths reaching 175 nm at 300 K.[35] However, our demonstration was with films prepared by sulfurization of evaporated Ag/Bi films, which had a discontinuous morphology that prevents their use in devices. Indeed, the majority of the current literature has focused on nanocrystals, since these can be spin-coated to form films free of pinholes, and which are sufficiently smooth to enable contact layers to be deposited by solution processing on top with continuous morphology. Future efforts therefore need to focus on developing processes that preserve this continuous morphology, while lowering grain boundary density and energetic disorder. Being able to increase $AgBiS_2$ absorber thicknesses to 400–500 nm, while maintaining efficient charge-carrier collection, is critical to reaching the high NIR spectral responses required to achieve tandem efficiencies approaching the 45.2% detailed balance limit for perovskite/$AgBiS_2$ 2J tandems, and 47.1% for 3J tandems.

Finally, the $V_{OC}$ achieved is comparable to the state-of-the-art for single-junction $AgBiS_2$ PVs (Fig. S14, SI), but these state-of-the-art devices have a $V_{OC}$ deficit >0.3 V below the radiative limited calculated from the absorption spectrum of $AgBiS_2$ (**Figure 2D** and Fig. S13, SI). This is in agreement with the low-intensity PL observed from these materials (**Figure 2E**). Possible causes of the $V_{OC}$ deficit may be from deep traps, interface recombination with the charge-transport layers, or energetic disorder. For example, our recent work on bulk films of $AgBiS_2$ led to reduced energetic disorder than nanocrystal films,[35] and could be a potential route to lowering the $V_{OC}$ deficit.

## Conclusion

We have established $AgBiS_2$ as a viable bottom cell material for perovskite-based tandem photovoltaics that has potential to enable higher efficiencies than Pb/Sn perovskite bottom cells, whilst also addressing their stability limitations. Detailed-balance and optical modeling showed that triple-junction

perovskite/perovskite/$AgBiS_2$ tandems have radiative limits of 47.5% with an 850 nm-thick $AgBiS_2$ absorber layer, surpassing the calculated 44.2% limit of equivalent devices using Pb/Sn perovskites as the bottom cell. Further, use of $AgBiS_2$ lowers the optimal bandgap of the mid- and top-cells; perovskites with <2 eV bandgap can be made with performance closer to their radiative limits. Realizing this potential experimentally required overcoming the severe limitation in preparing thick $AgBiS_2$ films with adequate NIR response. This limitation arises from the processing method conventionally used (solid-state ligand exchange), as well as the limited diffusion length of the absorber itself. We developed a solution-phase ligand exchange protocol with bromide/iodide co-passivation to produce colloidal nanocrystal inks suitable for one-step film deposition with controlled thicknesses up to 340 nm. The inclusion of bromide co-passivation reduced trap densities, thus reducing non-radiative recombination to enable improvements in open-circuit voltage and diffusion length. Our techno-economic analysis further highlights that the SPLE co-passivation method results in substantially reduced processing time and cost compared to SSLE, and we have shown that this approach is scalable from conventional lab-scale 1.2 × 1.2 $cm^2$ substrates to 3 × 3 $cm^2$ substrates, and is possibly compatible with large-area industrial solution deposition methods.

Using these mixed bromide/iodide co-passivated films as the NIR sub-cell, we made the first demonstration of perovskite/$AgBiS_2$ tandems, both four-terminal devices that reached 24% PCE, as well as monolithic two-terminal devices that successfully achieved voltage addition. We showed that the one-step SPLE method allows the $AgBiS_2$ PVs to be scalable from the 0.04 $cm^2$ area commonly used in the field thus far to the 0.25 $cm^2$ area expected for perovskite-based tandem development, representing an important maturation of this technology. Unencapsulated $AgBiS_2$ devices retained over 90% of their initial efficiency after three months under ambient shelf storage, underscoring the inherent stability advantage over oxidation-prone Pb/Sn bottom cells. In our roadmap to build upon this first demonstration, the critical bottleneck to address is to achieve $AgBiS_2$ films that avoid ultrafast carrier localization in order to increase diffusion lengths even further, while still maintaining continuous film morphology to retain high shunt resistances. Another area requiring further development is understanding where the open-circuit voltage loss originates in $AgBiS_2$ and finding routes to mitigate such losses. Realizing these improvements could allow $AgBiS_2$-based tandems to fulfil their potential as high efficiency, cost-effective, scalable and stable devices.

## Data availability

The data supporting the findings of this study is available within this paper are available upon reasonable request. Reused third-party data are cited in the reference list.

## Code availability

The detailed-balance, transfer-matrix and figure-generation code used in this study are available from the corresponding author upon reasonable request.

## Acknowledgements

The authors acknowledge funding from Nissan Technical Centre Europe and the Advanced Propulsion Centre (grant no. 10176695) through the DRIVE35 programme. The authors acknowledge the National Thin-Film Cluster Facility for Advanced Functional Materials at the Department of Physics, University of Oxford, funded by the Engineering and Physical Sciences Research Council (EPSRC; grant no. EP/M022900/1) and the Wolfson Foundation. The X-ray photoelectron spectroscopy measurements were supported by the National Natural Science Foundation of China (grant no. 52602273), the Collaborative Innovation Center of Suzhou Nano Science & Technology, the 111 Project, and the Joint International Research Laboratory of Carbon-Based Functional Materials and Devices. The spectroscopic ellipsometer was funded by the Engineering and Physical Sciences Research Council through the WAFT project (grant no. EP/M015173/1).

The authors thank Prof. Iain McCulloch for supporting the external quantum efficiency measurements, Prof. Moritz Riede for access to the spectroscopic ellipsometer, and Dr. Jin Yao for performing IZO and Ag thin-film depositions at the National Thin-Film Cluster Facility for Advanced Functional Materials.

R. L. Z. H., Yo. W. and S. D. acknowledge funding from UK Research and Innovation through a Frontier Grant (grant no. EP/X029900/1). Yi. W. acknowledges funding from The Hilary and Galen Weston Foundation at Merton College, University of Oxford, and the Department of Chemistry, University of Oxford. R. F. and H. J. S. acknowledge funding from the EPSRC-funded Programme Grant, titled Advanced Device Concepts for Next-Generation Photovoltaics (no. EP/X038777/1). R. S. B and J. O. were supported by the EPSRC (no. EP/X037169/1, UKRI4166), and by the Leverhulme Trust's Philip Leverhulme Prize in Engineering (no. PLP-2022-154). R. L. Z. H. acknowledges funding from the Royal Academy of Engineering and the Science and Technology Facilities Council through the Research Chair scheme (grant no. RCSRF2324-18-68).

## Author contributions

R. L. Z. H. and Yo. W. conceived of this project. S. D. and Yo. W. developed the $AgBiS_2$ nanocrystal synthesis and ligand-exchange processes, fabricated and characterized the $AgBiS_2$ photovoltaic devices, and analyzed the device performance, supervised by R. L. Z. H. Hae-Jun Seok and M. H. conducted perovskite and tandem device fabrication and characterization, supervised by R L. Z. H. E. N. and D. Y. C.

assisted with device fabrication and optimization, with D. Y. C. co-supervised by L. W. G. C. developed and optimized the device layout for reproducible tandem fabrication. R. F. and S. S. developed and optimized the IZO deposition process to produce high-quality transparent electrodes, supervised by H. J. S. J. O. helped with measuring the electronical properties of $H:In_2O_3$, supervised by R. S. B. Yi. W. and J. M. W. performed and analyzed the OPTP measurements. X. G. performed EQE characterization. K. T. and D. I. contributed to project discussions and supported funding acquisition. A. R. conducted and analyzed PDS measurements, supervised by A. S. Y. Y. and Q. S. performed and analyzed PL measurements. H. K. P. performed AFM measurements, supervised by J.-S. K. S. D. and Yo. W. analyzed the data and prepared the manuscript with R. L. Z. H. All authors discussed the results and contributed to writing and editing the manuscript.

## Competing interests

A patent application related to this work was filed on 10/5/26 (GB2623364.3), for which R.L.Z.H., S.D., Yo.W and Yi.W. are listed as inventors. R.L.Z.H. consults for NanoPrint Innovations Ltd, a startup commercializing spatial atomic layer deposition (SALD) for applications that include tandem photovoltaics. D.I. and K.T. are employees of Nissan, and are part of an Innovate UK funded consortium that has the development of electric vehicles with tandem photovoltaics integrated as one of its aims. H.J.S. is co-founder and CSO of Oxford Photovoltaics Ltd., which is commercialising perovskite/silicon tandems. G.C. is co-founder and co-CEO of Ark Metrica Ltd.